\documentclass[12pt,onecolumn,a4paper,superscriptaddress,floatfix]{revtex4}
\usepackage{graphicx}

\begin{document}

\title{Macroscopic polarization entanglement \\ 
       and loophole-free Bell inequality test}

\author{M.~Stobi\'nska}
\email{magda.stobinska@mpl.mpg.de}
\affiliation{Institute for Theoretical Physics II,
  Erlangen-N\"urnberg University, Erlangen, Germany}
\affiliation{Max Planck Institute for the Science of Light,
  Erlangen, Germany}
\author{P.~Horodecki}
\affiliation{Faculty of Applied Physics and Mathematics,
  Gda{\'n}sk University of Technology, Gda{\'n}sk, Poland}
\affiliation{National Quantum Information Center of Gda\'nsk, Sopot, Poland}
\author{A.~Buraczewski}
\affiliation{Faculty of Electronics and Information Technology,
  Warsaw University of Technology, Warsaw, Poland}
\author{R.~W.~Chhajlany}
\affiliation{Faculty of Applied Physics and Mathematics,
  Gda{\'n}sk University of Technology, Gda{\'n}sk, Poland}
\affiliation{National Quantum Information Center of Gda\'nsk, Sopot, Poland}
\affiliation{Faculty of Physics, Adam Mickiewicz University,
  Pozna\'n, Poland}
\author{R.~Horodecki}
\affiliation{National Quantum Information Center of Gda\'nsk, Sopot, Poland}
\affiliation{Institute of Theoretical Physics and Astrophysics,
  University of Gda\'nsk, Gda\'nsk, Poland}
\author{G.~Leuchs}
\affiliation{Max Planck Institute for the Science of Light,
  Erlangen, Germany}
\affiliation{Institute for Optics, Information and Photonics,
  Erlangen-N\"urnberg University, Erlangen, Germany}


\maketitle

{\bf Quantum entanglement \cite{EPR,Schroedinger} revealed the
  inconsistency between the classical and the quantum laws governing
  the living and inanimate matter \cite{Zukowski,Norsen}. Quantum
  mechanical predictions contradict local realistic theories
  \cite{EPR} leading to a violation of Bell inequalities
  \cite{Bell,CHSH} by entangled states. All experiments confirming the
  violation suffered from loopholes
  \cite{Aspect1982,Weihs1998,Rowe2001,Matsukevich2008}, a fundamental
  problem in modern physics
  \cite{Garcia2004,Cabello2008,Vertesi2010}. Detection loopholes
  result from data postselection due to inefficient photodetection of
  single quanta.  Large quantum systems though difficult to produce,
  bring us closer to complex biological organisms allowing for testing
  their quantum nature
  \cite{Arndt2009,Romero2010,Mohan2010,Collini2010}.  Here we show
  that loophole-free Bell tests are possible within the current
  technology using multi-photon entanglement \cite{DeMartini-PRL} and
  linear optics. A preselection protocol prepares the macroscopic in
  photon number entanglement ($10^3$ photons at least) in advance,
  making the postselection unnecessary. Complete loss of the state is
  impossible and dark counts are negligible.  Fast switching
  measurements close the locality loophole. Macroscopic preselected
  states find application in creating quantum superpositions of living
  organisms and their manipulation \cite{Romero2010}.}

\begin{figure}[h]
\begin{center}
  \raisebox{7cm}{\hbox to 0pt{\hss (a)}}
  \includegraphics[height=6cm]{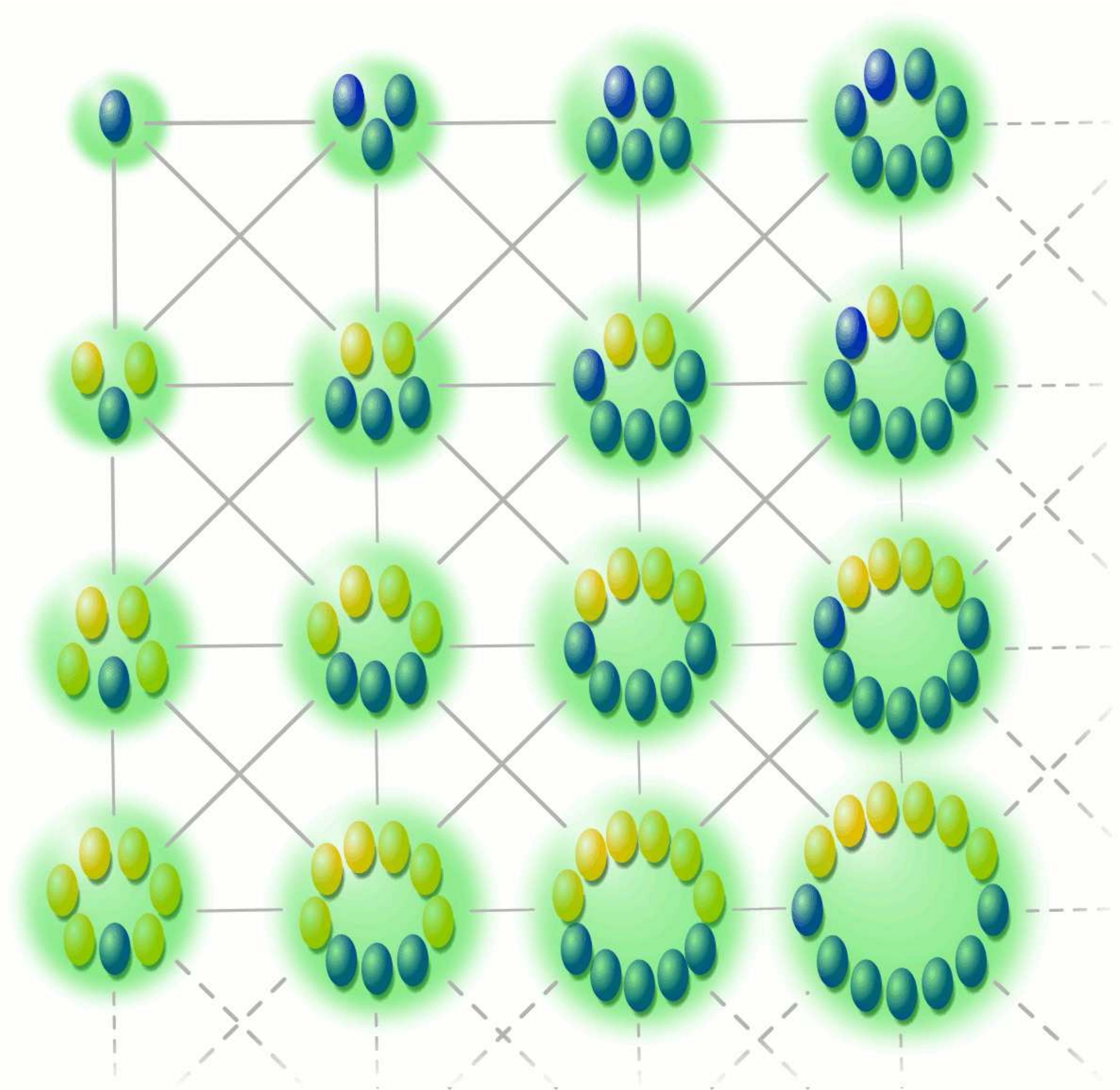}
  \hskip1cm
  \includegraphics[height=6.5cm]{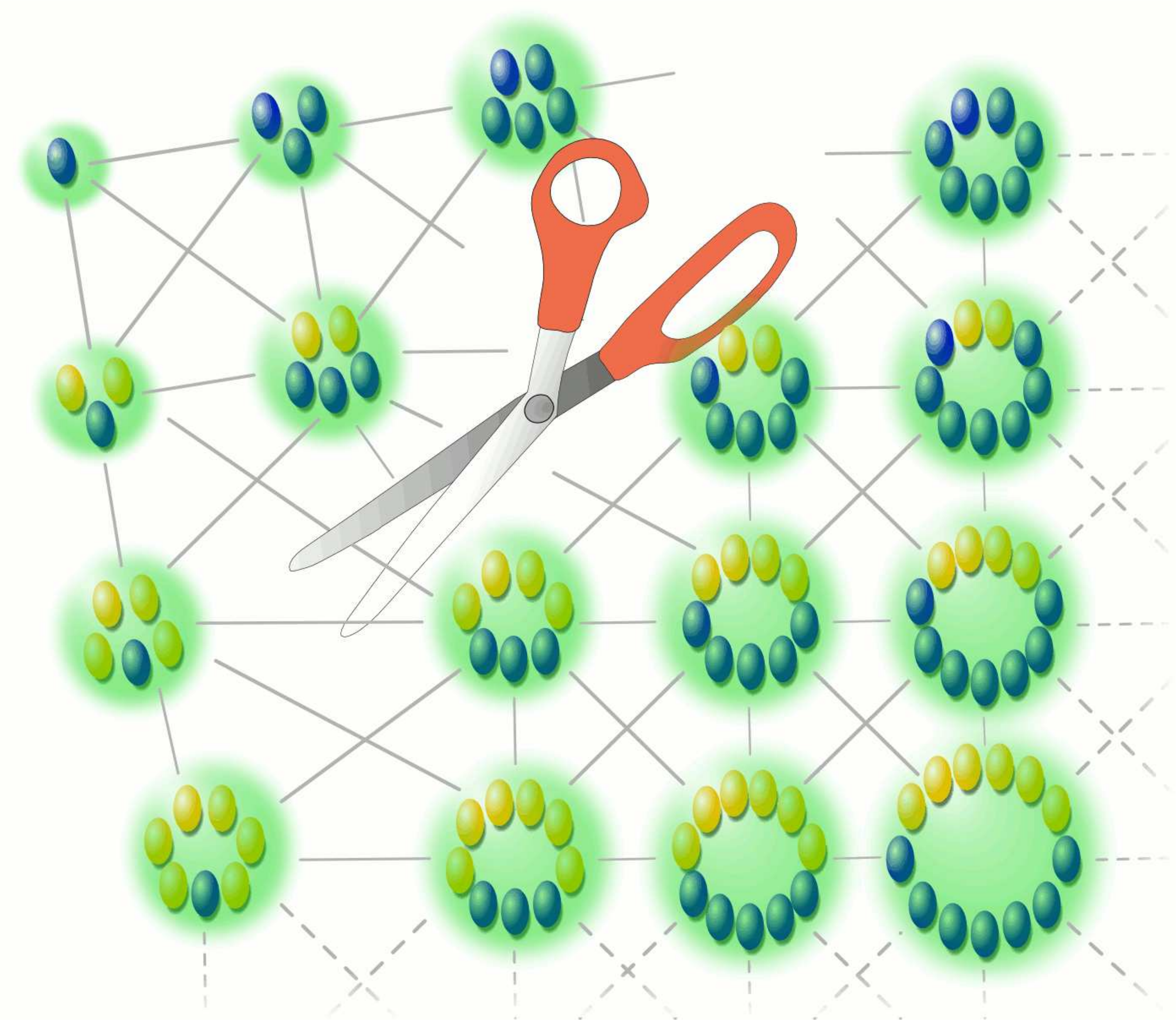}
  \raisebox{7cm}{\hbox to 0pt{\hss (b)}}
\end{center}
\caption{(a) Multi-photon state $|\Phi\rangle$ consists of all
  combinations of odd photon numbers in $\varphi$ (blue ellipses) and
  even photon numbers in $\varphi^{\perp}$ (yellow ellipses)
  polarization. (b) Quantum scissors (preselection protocol) cut-off
  the small photon numbers.}
\label{multiphoton}
\end{figure}

The multi-photon polarization entanglement has recently been
demonstrated \cite{DeMartini-PRL,DeMartini-single,Masha}.  First, a
two--mode pair of linearly polarized photons is created in a standard
singlet through parametric down conversion (PDC).  We describe the PDC
process in terms of polarization states lying in the equatorial plane,
$|1_{\varphi}\rangle = 1/\sqrt{2}(|1_H, 0_V\rangle + e^{i
  \varphi}|0_H, 1_V\rangle)$ and its orthogonal counterpart
$|1_{\varphi^{\perp}}\rangle $, parametrized by the polar angle
$\varphi$, where $|n_H, m_V\rangle$ denotes $n$ ($m$) photons
polarized horizontally (vertically). The singlet takes the form
\begin{equation}
\frac{1}{\sqrt{2}}(|1_{\varphi}\rangle_A
|1_{\varphi^{\perp}}\rangle_B -|1_{\varphi^{\perp}}\rangle_A
|1_{\varphi}\rangle_B).
\label{singlet}
\end{equation}
Next, the population of each PDC outcoming spatial mode ($A$ and $B$)
can be independently phase sensitive amplified to create a
multi-photon state by passing the appropriate photon through an
intensely pumped high gain $g$ nonlinear medium. We denote amplified
$|1_{\varphi }\rangle$ and $|1_{\varphi^{\perp}}\rangle$ multi-photon
states for a fixed $\varphi$ as $|\Phi\rangle$ and
$|\Phi_{\perp}\rangle$, respectively. Those states reveal interesting
interplay between polarization and photon number degrees of freedom:
$|\Phi\rangle$ consists of all combinations of odd photon numbers
$(1,3,5,...)$ in $\varphi$ and even photon numbers $(0,2,4,...)$ in
$\varphi^{\perp}$ polarization whereas $|\Phi_{\perp}\rangle$ consists
of all combinations of even photon numbers in $\varphi$ and odd photon
numbers in $\varphi^{\perp}$ polarization (see
Fig.\ref{multiphoton}(a)).  Due to different parity of photon numbers
in $\varphi$ and $\varphi^{\perp}$ polarizations these states are
orthogonal.  In the experiment, they contained up to $4m=10^{4}$
photons on average, where $m=\mathrm{sh}^2g$. However, small as well
as large photon number components contribute to them.  Amplification
of one singlet mode e.g. $B$, leads to a ``micro-macro'' singlet
$1/\sqrt{2}(|1_{\varphi}\rangle_A |\Phi_{\perp}\rangle_B -
|1_{\varphi^{\perp}}\rangle_A |\Phi\rangle_B)$.  If both modes were
amplified, a ``macro-macro'' entangled state $1/\sqrt{2}(
|\Phi\rangle_A |\Phi_{\perp}\rangle_B - |\Phi_{\perp}\rangle_A
|\Phi\rangle_B)$ would be produced.  The Bell test with these singlets
would not be practical since multi-photon states $|\Phi\rangle$ and
$|\Phi_{\perp}\rangle$ have to be fully distinguishable. Although the
probability distributions
$Q_{\Phi}(n_{\varphi},n_{\varphi^{\perp}})=|\langle
n_{\varphi},n_{\varphi^{\perp}}|\Phi\rangle|^{2}$ and
$Q_{\Phi_{\perp}}(n_{\varphi},n_{\varphi^{\perp}})$ do not overlap on
the single photon scale (see Fig.\ref{fig1}(a) and (b)) there are no
detectors allowing parity measurements for intense beams. An effective
overlap (resulting from small photon numbers) of order of $10^{-1}$ is
measured as if $Q$--functions were continuous \cite{DeMartini-PRL}(see
Fig.\ref{fig1}(c)).

\begin{figure}[p]
\begin{center}
  \raisebox{7.5cm}{\hbox to 0pt{\hss (a)}}
  \includegraphics[height=7cm]{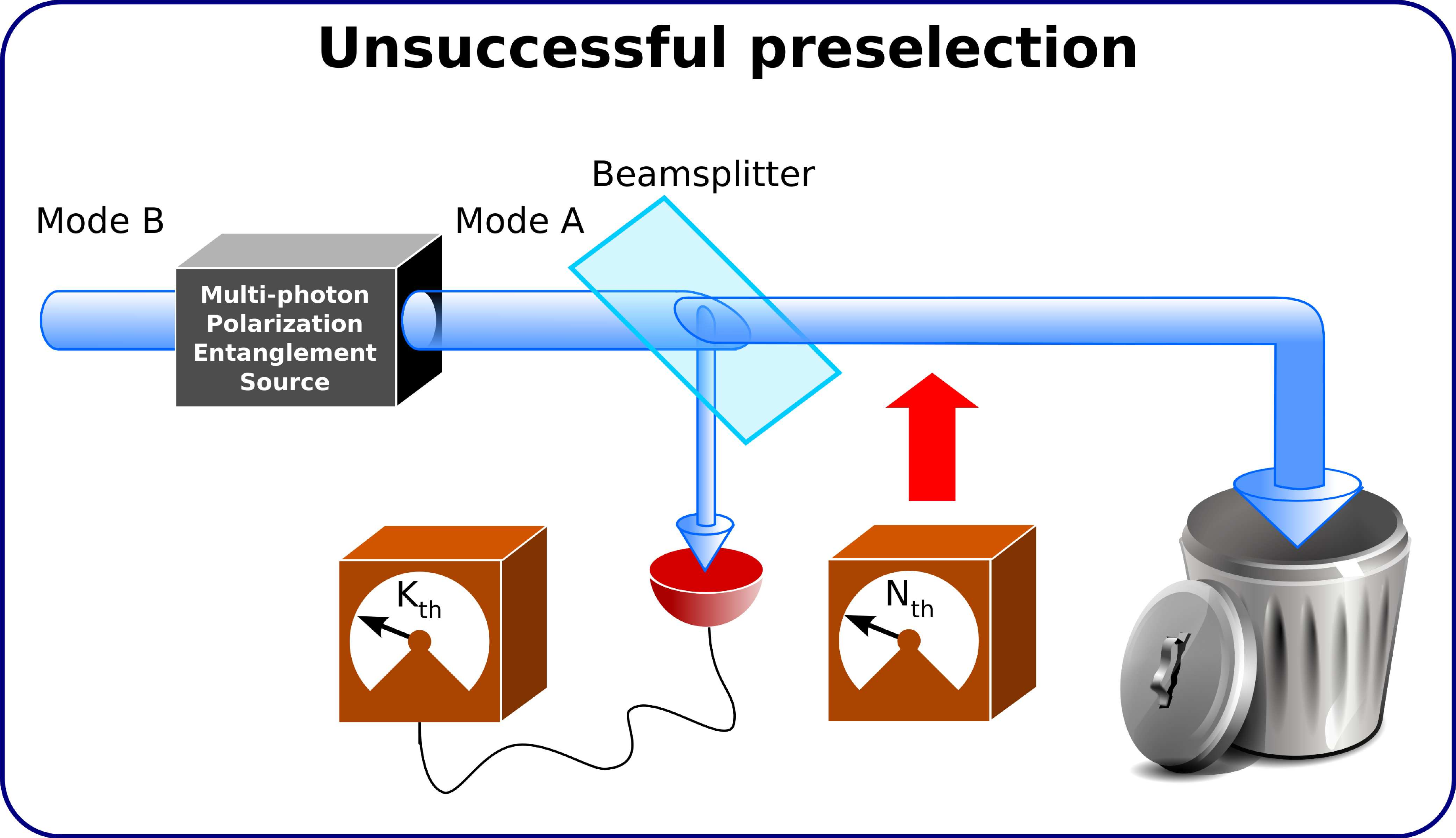}
  \\
  \raisebox{7.5cm}{\hbox to 0pt{\hss (b)}}
  \includegraphics[height=7cm]{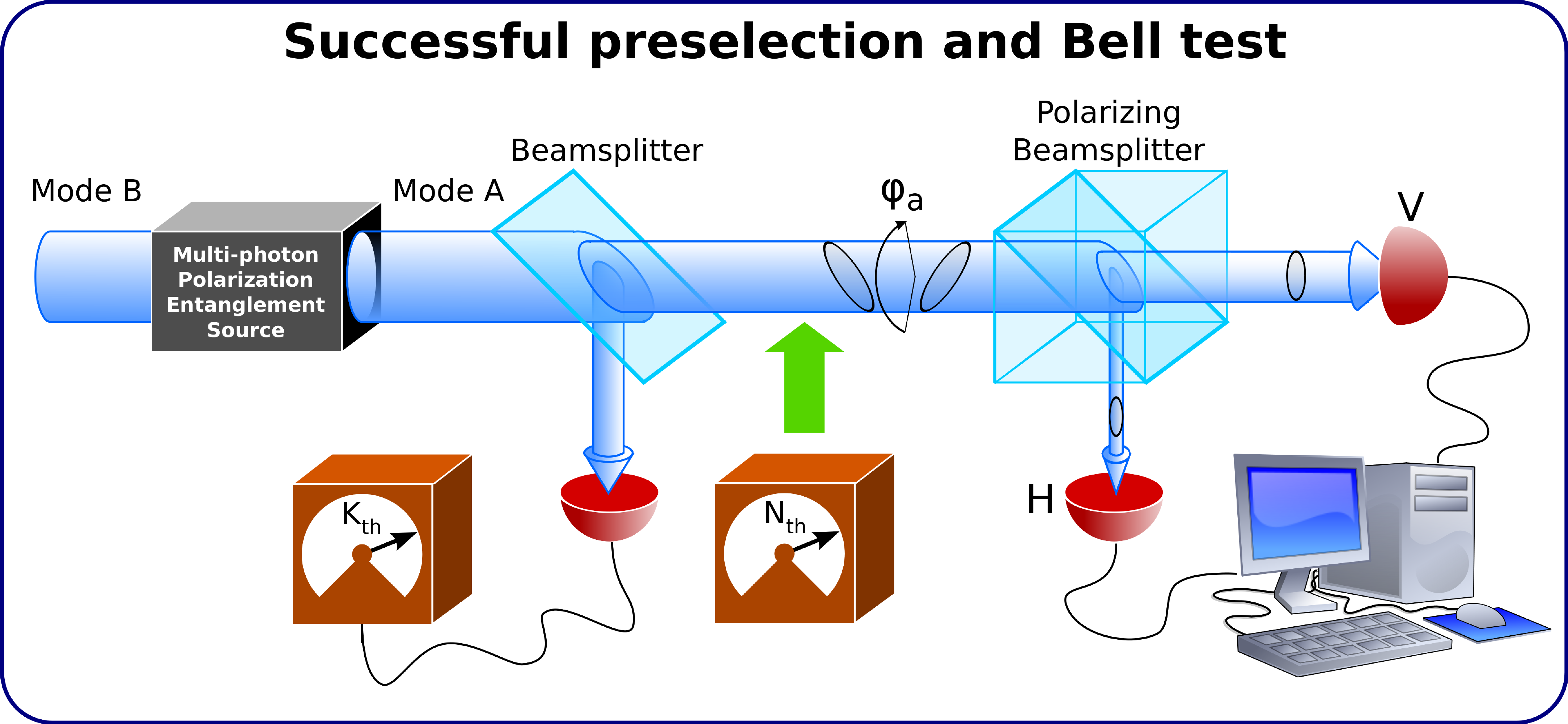}
\end{center}
\caption{The schematic for conditional preselection protocol (only
  mode $A$ is shown).  A multi-photon polarization entanglement source
  is followed by beamsplitters BS, phaseshifters PS and polarizing
  beamsplitters PBS.  The Bell test is performed using preselected
  ``macro-macro'' polarization singlet only if the detectors measuring
  the reflected beams report photon numbers greater than the threshold
  $K_{th}$ (b). Otherwise the state is rejected (a).}
\label{fig3}
\end{figure}

\begin{figure}[h]
\begin{center}
  \hskip-1cm
  \includegraphics[height=7cm]{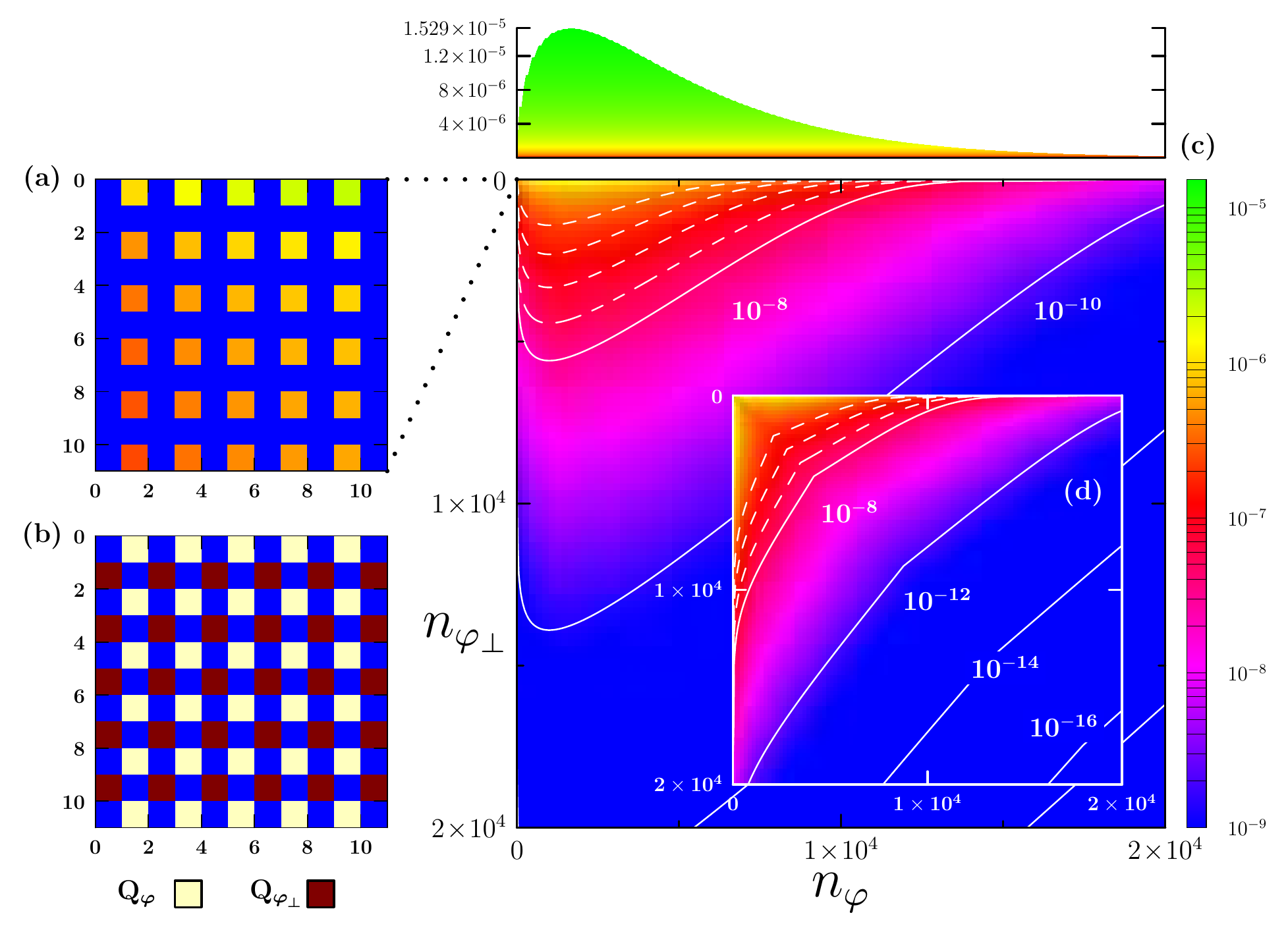}
  \hskip0.2cm
  \includegraphics[height=7cm]{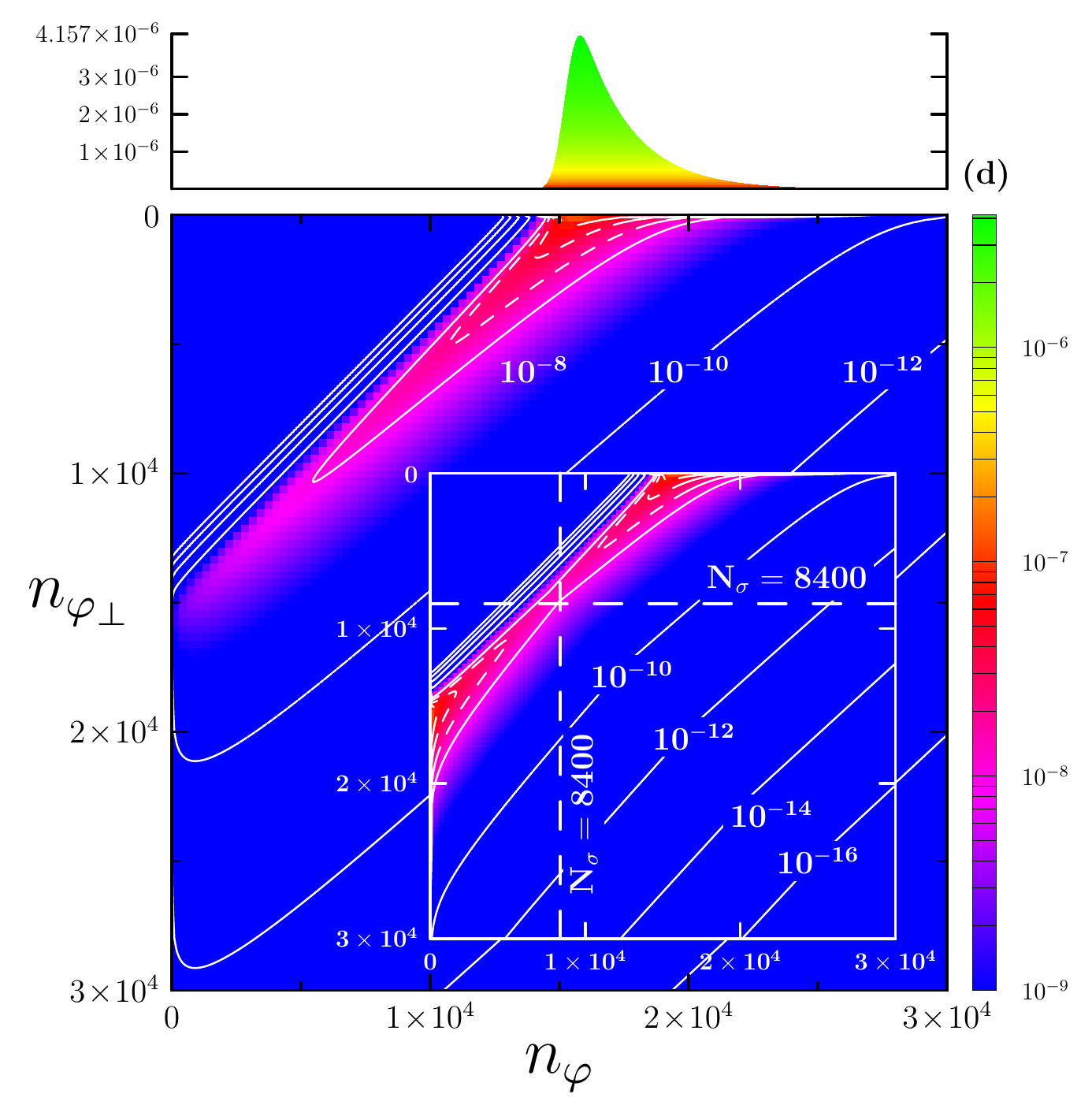}
  \hskip-3cm
\end{center}
\caption{The $Q_{\Phi}(n_{\varphi},n_{\varphi^{\perp}})$ and
  $Q_{\Phi_{\perp}}(n_{\varphi},n_{\varphi^{\perp}})$ distributions
  for $|\Phi\rangle$ and $|\Phi_{\perp}\rangle$ with $m\!=\!10^3$,
  respectively.  (a) before preselection $Q_{\Phi}$ is discrete on the
  single photon scale but (c) continuous on the macroscopic scale of
  detection. (b) the distributions $Q_{\Phi}$ and $Q_{\Phi_{\perp}}$
  do not overlap on the single photon scale however, (d) huge
  effective overlap is measured. (e) $Q_{\Phi}$ is shifted to high
  photon number region after preselection with threshold
  $K_{th}\!=\!1700$. (f) the distributions $Q_{\Phi}$ and
  $Q_{\Phi_{\perp}}$ for the same threshold.  $Q_{\Phi}$
  ($Q_{\Phi_{\perp}}$) is highly concentrated around the $n_{\varphi}$
  ($n_{\varphi^{\perp}}$) axis with a shadow constrained approximately
  within the regions $\{ n_{\varphi^{\perp}} \!\le \!N_{\sigma}$,
  $n_{\varphi} \!>\!  N_{\sigma} \}$ ($\{ n_{\varphi} \! \le \!
  N_{\sigma}$, $n_{\varphi^{\perp}}\! >\! N_{\sigma} \}$), where
  $N_{\sigma}\!=\!8400$.}
\label{fig1}
\end{figure}

Our proposal does not rely on the orthogonality of the odd and even
Fock states. Here, we propose a new preselection protocol which shifts
the entanglement to high photon-numbers and makes the $Q$--functions
unambiguously distinguishable for detection.  It acts as quantum
scissors which cut-off the small photon number contributions in
multi-photon states and thus produce a genuine macroscopic state
useful for a Bell test (see Fig.\ref{multiphoton}(b)).  It rejects
states for which $n_{\varphi} + n_{\varphi^{\perp}} \le N_{th}$ where
$N_{th}$ is a certain (approximate) threshold for the acceptable
smallest number of photons in each preselected multi-photon state, see
Fig.~\ref{fig3}(a). The above condition imposes a constraint only on
the sum of the two photon numbers but gives no information about the
polarization components.  After preselection the odd-even structure of
the $Q$--functions is lost. Ideally, the process is carried out by the
projector
\begin{equation}
{\cal P}_{n_{\varphi} + n_{\varphi^{\perp}} > N_{th}} \equiv
\!\!\!\!\! \sum_{n_{\varphi},n_{\varphi^{\perp}}: n_{\varphi} +
  n_{\varphi^{\perp}} > N_{th}} \!\!\! \!\!\!\!\!\!\!\!
\!\!\!\!\!|n_{\varphi} \rangle \langle n_{\varphi}| \otimes
|n_{\varphi^{\perp}} \rangle \langle n_{\varphi^{\perp}}|.
\end{equation}
This however cannot be perfectly implemented with current technology,
though an arbitrarily good approximation may be obtained for high
average photon numbers.  If $N_{th}$ is high enough, the preselected
macroscopic entangled output state approaches a macroscopic singlet
state.

We will focus on ``macro-macro'' entanglement, though the argument is
adoptable to the ``micro-macro'' case.  

Each multi-photon state from a multi-photon polarization entanglement
source is passed through a preselecting unbalanced beamsplitter with a
small reflectivity $R$, e.g. $R\!=\!10\%$ and the reflected beam
intensity is measured.  The transmitted beam is accepted for Bell test
if the number of reflected photons is greater than an appropriately
chosen threshold $K_{th}$, Fig.~\ref{fig3}(b). A large reflected
intensity $k$ means that a very large number, at least $N_{th}$, of
photons are transmitted in the process given the strong bias of the
beamsplitter. Indeed, reflection of more than $2R\!=\!20\%$ of the
impinging photons has negligible probability, implying that at least
$80\%$ of the photons must have been transmitted.  To get a lower
bound on the number of transmitted photons $n$, we assume the
reflected number of photons to constitute the mentioned $20\%$ of
input photons and hence infer from the measured intensity $k\ge
K_{th}$ (or $k<K_{th}$) if $n \ge N_{th}$ (or $n<N_{th}$).  The
transmitted beam is not in pure state for an arbitrary $K_{th}$ (since
a beamsplitter entangles reflected and transmitted beams) however, it
is approximately in pure (projected) state for large $K_{th}$.  The
scheme works well for highly populated input states. The $Q$-functions
for the preselected macroscopic states with $m\!=\!10^3$ and
$K_{th}\!=\!1700$ obtained with the probability of success
$p\!=\!2\cdot 10^{-3}$ are presented in Fig.~\ref{fig1}(b).  They are
practically disjoint with overlap equal to $8 \cdot 10^{-5}$.  In
principle, one could preselect multi-photon states with a smaller
photon number but the probability drops dramatically.

After successful preselection in both modes, the output 
entangled state can be used for a Bell test.  The test with
macroscopic singlets is highly desirable, as the probability of losing
the state is negligible and dark counts are easy to notice.  Similarly,
small single photon detection efficiency is not a limiting factor even
for a preselected ``micro-macro'' entangled state \cite{BrunnerEtAl}.
One may choose the basic observable in the form of a three-output
intensity measurement
\begin{equation}
\mathcal{A}(\varphi )={\cal P}_{n_{\varphi} \le N_{\sigma}} \otimes
        {\cal P}_{n_{\varphi^{\perp}} > N_{\sigma}} - {\cal
          P}_{n_{\varphi} > N_{\sigma}} \otimes {\cal
          P}_{n_{\varphi^{\perp}} \le N_{\sigma}}.
\label{obs1}
\end{equation}
The eigenvalue -1(+1) corresponds approximately to a measurement of
preselected state $|\Phi\rangle (|\Phi_{\perp}\rangle )$, {\it i.e.}
the upper (lower) off--diagonal quadratures in Fig.\ref{fig1}(b).  The
observable suggests the proper value of the $N_{\sigma }$ parameter,
which can be set at the detectors, for the examined preselected states
for a given $K_\mathrm{th}$. It allows to maximally profit from the
disjointness of the preselected states while minimizing the discarded
part of the $Q$--functions. This feature weighs heavily on the
correlation between the two macroscopic states. In short, $N_{\sigma
}$ defines the maximal number of $\varphi^{\perp}$ ($\varphi$) -
polarized photons in the preselected states.  For the considered case
$N_{\sigma}=8400$. The observable also accounts for violation of the
filtering condition (measurement of either of the two diagonal
quadratures in Fig.\ref{fig1}(b)) due to imperfections in the
preselection process yielding 0 in such circumstances. These results
are inconclusive for Bell test and they contribute to the loophole
(see Methods). Alternatively, one may choose a binary observable on
only one of the polarization modes of preselected state
\begin{equation}
\bar{\mathcal{A}}(\varphi)= [ {\cal P}_{n_{\varphi} <
    N_{\sigma}} - {\cal P}_{n_{\varphi} \geq N_{\sigma}} ] \otimes I.
\label{obs2}
\end{equation} 
 
The Bell test is carried out using a Bell-CHSH observable
\begin{eqnarray}
\mathcal{B} &=& \mathcal{O}(\varphi_{a}) \otimes
\mathcal{O}(\varphi_{b}) + \mathcal{O}(\varphi_{a}) \otimes
\mathcal{O}(\varphi_{b'}) \nonumber\\ &+& \mathcal{O}(\varphi_{a'})
\otimes \mathcal{O}(\varphi_{b})- \mathcal{O}(\varphi_{a'}) \otimes
\mathcal{O}(\varphi_{b'}),
\label{Bell-observable}
\end{eqnarray}  
where ${\cal O}(\varphi )\!=\!{\cal A}(\varphi ), \bar{{\cal
    A}}(\varphi ) $ depending on the choice of the basic observable.
Observable ${\cal O}$ is measured in the following steps: (i) the
polarization of each macroscopic state is first independently rotated
about the axis perpendicular to the equatorial plane of polarization
through the angle $\varphi$ using a Babinet-Soleil phase shifter (PS),
(ii) followed by intensity measurements of the two reference
polarization modes $\varphi \!=\!0, \pi $. The angles are chosen as
$\varphi_{a}\!=\!0$, $\varphi_{a'}\!=\!\frac{\pi}{4}$,
$\varphi_{b}\!=\!\frac{\pi}{8}$,
$\varphi_{b'}\!=\!\frac{3}{8}\pi$. Bell inequality violation
corresponds to $2 \!<\! |\langle \mathcal{B} \rangle| \!\leq \! 2
\sqrt{2}$.

\begin{figure}[h]
\begin{center}
  \includegraphics[height=6cm]{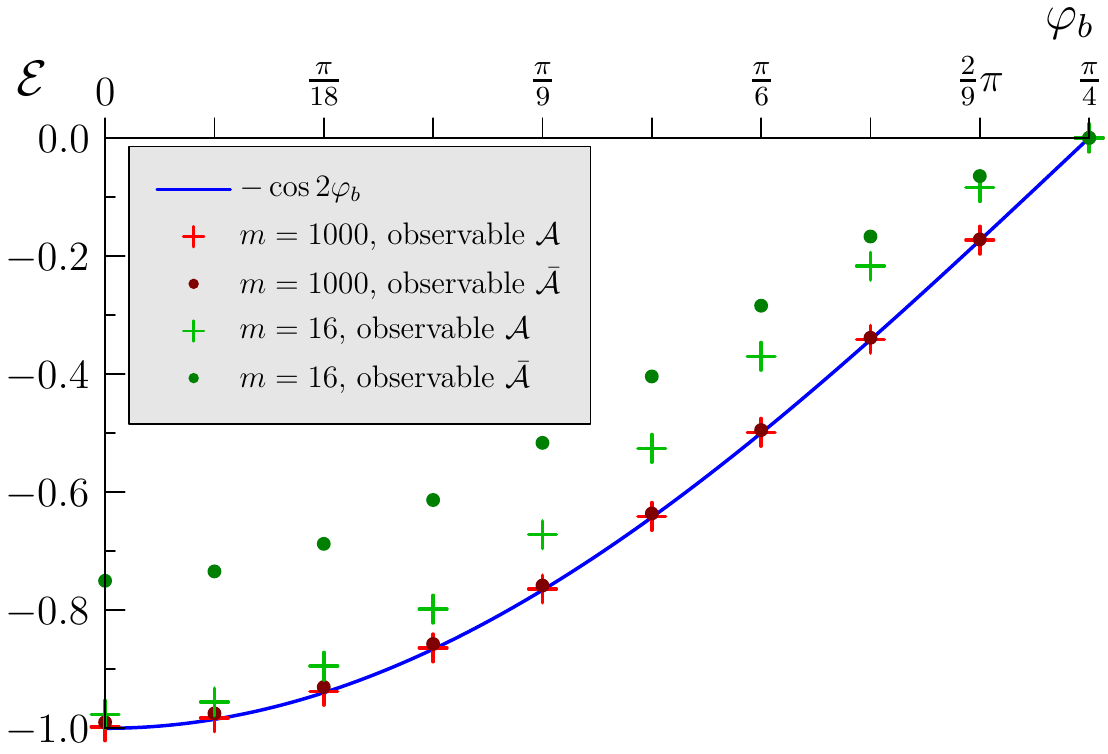}
\end{center}
\caption{Correlation function $\mathcal{E}=\langle \mathcal{O}(0)
  \otimes \mathcal{O}(\varphi_{b}) \rangle$ evaluated for preselected
  ``macro-macro'' singlet with $m\!=\!16$ and $K_{th}\!=\!70$ -- green
  points and $m\!=\!10^3$ and $K_{th}\!=\!1700$ -- red points.}
\label{fig4}
\end{figure}

The correlations $\langle \mathcal{O}(0) \otimes
\mathcal{O}(\varphi_{b}) \rangle$ for states with mean number of
photons $m\!=\!16$ with $K_{th}\!=\!70$ and $m\!=\!10^3$ with
$K_{th}\!=\!1700$ are depicted in Fig.\ref{fig4}.  For $m\!=\!10^3$
and perfectly aligned observables, $\langle \mathcal{O}(0) \otimes
\mathcal{O}(\varphi_{b}) \rangle\!=\!-1$ revealing perfect
singlet-like correlations in the state.  However, for $m\!=\!16$
observable $\bar{{\cal A}}(\varphi )$ reveals worse correlation than
${\cal A}(\varphi )$.  Further testimony to the quality of state is
provided by recalling that the angular dependence of the correlation
in a singlet is given by $(-\cos 2\varphi_b)$. This dependence
characterizes the preselected state accurately with rise in the mean
number of photons.  Also discrepancy between the values of the two
observables decreases. For parameters shown in Fig.\ref{fig1}, the
best and almost maximal Bell inequality violation $\left\langle
\mathcal{B} \right \rangle_{\mathcal{A}}\!\!=\!\!-2.82$ and
$\left\langle
\mathcal{B}\right\rangle_{\bar{\mathcal{A}}}\!\!=\!\!-2.80$ is indeed
obtained for $N_{\sigma }\!=\!8400$ for both observables with
probability $p^2\!=\!3.9 \times 10^{-6}$.  The observable
$\mathcal{A}$ becomes loophole-free $\mathcal{L} \!\!=\!\!
0.003$. The results corresponding to $\bar{\mathcal{A}}$ are less
perfect since it better witnesses deviations of the input state
correlations from the singlet-like. The differences in results become
more pronounced and lead to contradicting conclusions for non-optimal
thresholds.  Higher probability rates for the Bell inequality
violation can be achieved for lower values of thresholds
e.g. $p^2\!\!=\!\!1.7 \times 10^{-3}$ is obtained for
$K_{th}\!\!=\!\!1000$ and $N_{\sigma}\!=\!5200$. In this case
$\left\langle \mathcal{B} \right \rangle_{\mathcal{A}}\!\!=\!\!-2.78$
with $\mathcal{L} \!\!=\!\!0.02$ and $\left\langle
\mathcal{B}\right\rangle_{\bar{\mathcal{A}}}\!\!=\!\!-2.65$ is
obtained.  In principle, Bell inequality violation is possible even
for $m\!=\!16$ however, it is not practical due to low probability of
success. For further details see Methods section.

Importantly, our analysis shows that the scheme will work even in
presence of losses or equivalently inefficient preselection. The
detectors measure the threshold value $K_{th}^{meas}$ with some
uncertainty, thus its value is usually known up to $150$
photons. Therefore, our input state should be rather considered as a
mixture of preselected macroscopic entangled states for $K_{th}
\!\in \!\langle K_{th}^{meas}\!-\!150, K_{th}^{meas}\!+\!150 \rangle$
than a single state for a given $K_{th}^{meas}$. However, for large
$m$ the higher values of $N_{\sigma}$, the broader window of $K_{th}$
for which the preselected state reveals near perfect correlations and
Bell inequality violation for the fixed measurement settings.  For
example, for $N_{\sigma}\! \in \!\langle 8200, 8600\rangle$ we have
the window of $K_{th}\! \in \!\langle 1400, 2400\rangle$.  This means
that mixedness of the input state will not destroy Bell inequality
violation since every term in the mixture leads to violation for the
same set of angles.  Other imperfections caused by absorption or
dephasing of photons in the transmitted beam are negligible at short
distances.

Our results find immediate application for testing the quantum nature 
of biological systems. Preselected macroscopic states are useful for 
creation of quantum non-gaussian superpositions of position and momentum 
of optically trapped species in a cavity via teleportation protocols \cite{Romero2010,Ashkin1987}. 
Creation of virus-light "micro-macro" singlet state via entanglement swapping 
will allow for manipulations on virus degrees of freedom in the strong 
coupling limit. Alternatively, the measurement performed on the light 
beam my allow to trace the virus position in the cavity.

In conclusion, we have shown the preselected, conditionally generated
macroscopic entanglement to be a powerful resource for loophole-free
Bell inequality testing within the current technology. These results
are a proof-of-principle of the accessibility of such resources in
real experimental situations. They find application in interdisciplinary 
research on testing of the quantum nature of biological systems as well 
as could be useful for quantum teleportation, security verification in 
quantum cryptology, controlled experimental analysis of quantum-to-classical 
transition phenomena \cite{Zurek} and experimental Bell tests using detectors 
with various detection profiles such as human eyes \cite{Gisin-eye}.

\section*{Methods} 

(i) The multi-photon states resulting from single photon amplification
are as follows
\begin{eqnarray}
|1_{\varphi }\rangle \rightarrow |\Phi\rangle \!\!\!&=&
\!\!\!\sum_{i,j=0}^{\infty} \!\gamma_{ij}
\big|(2i+1)_{\varphi},(2j)_{\varphi^{\perp}}\rangle,
\label{macro-qubits}
\\|1_{\varphi^{\perp}}\rangle \rightarrow |\Phi_{\perp}\rangle
\!\!\!&=&\!\!\! \sum_{i,j=0}^{\infty} \!\gamma_{ij}
\big|(2j)_{\varphi},(2i+1))_{\varphi^{\perp}}\rangle, \nonumber
\end{eqnarray}
with $\gamma_{ij}=C^{-2}(-\frac{\Gamma}{2})^i(\frac{\Gamma}{2})^j
\frac{\sqrt{(1+2i)!(2j)!}}{i!j!}$, $C = \mathrm{cosh} (g)$, $\Gamma=
\mathrm{tanh} (g)$.

(ii) The beamsplitter (BS) operation $U_{BS}|0,N\rangle$
$\!=\!\sum_{k=0}^N c_k^{(N)} |k,N-k\rangle$ leads to the following
probability amplitudes $c_k^{(N)} \!=\!{N \choose k}^{1/2} (R^k
\left(1-R\right)^{N-k})^{1/2}$ for reflection of $k$ and transmitting
of $N-k$ photons. The BS is polarization independent.  The BS
preselected multi-photon states are as follows
\begin{eqnarray}
|\tilde{\Phi}\rangle_{tr} \!\!&=&\!\! \mathcal{N}\!\!
\sum_{i,j=0}^{\infty} \!\!\gamma_{ij}\!\! \sum_{n=0}^{2i+1}
\sum_{m=m^{*}}^{2j} c_n^{(2i+1)} c_m^{(2j)}
|(n)_{\varphi},(m)_{\varphi^{\perp}} \rangle_r \nonumber\\ &{}&
\otimes |(2i+1-n)_{\varphi},(2j-m)_{\varphi^{\perp}}
\rangle_t,\nonumber\\ |\tilde{\Phi}_{\perp}\rangle_{tr} \!\!&=&\!\!
\mathcal{N}\!\! \sum_{i,j=0}^{\infty} \!\!\gamma_{ij} \!\!
\sum_{n=0}^{2j} \sum_{m=m^{*}}^{2i+1} c_n^{(2j)} c_m^{(2i+1)}
|(n)_{\varphi},(m)_{\varphi^{\perp}} \rangle_r \nonumber\\ &{}&
\otimes |(2j-n)_{\varphi},(2i+1-m)_{\varphi^{\perp}}
\rangle_t\nonumber,
\end{eqnarray}
where $r$ and $t$ denote the reflected and transmitted beams
respectively, the normalization constant equals $\mathcal{N} \!=\!
\sum_{i,j=0}^{\infty} |\gamma_{ij}|^2 \sum_{n=0}^{2i+1}
\!\sum_{m=m^{*}}^{2j} |c_n^{(2i+1)}| |c_m^{(2j)}|^2$ and
$m^{*}\!=\!{\rm max}(0,K_{th}-n)$. The preselected macroscopic density
operators (after the measurement on the reflected beam) for the Bell
test are obtained by tracing over the reflected beam
$\hat{\rho}(\tilde{\Phi}) = \mathrm{Tr}_r\{
|\tilde{\Phi}\rangle_{tr}\langle \tilde{\Phi}|\}$,
$\hat{\rho}(\tilde{\Phi}_{\perp}) = \mathrm{Tr}_r\{
|\tilde{\Phi}_{\perp}\rangle_{tr}\langle \tilde{\Phi}_{\perp}|\}$. The
``macro-macro'' singlet state is given by
\begin{eqnarray}
\hat{\rho} &=& 1/2 \times \mathrm{Tr}_r\left\{ \left(
|\tilde{\Phi}\rangle_{tr} |\tilde{\Phi}_{\perp}\rangle_{tr} -
|\tilde{\Phi}_{\perp}\rangle_{tr} |\tilde{\Phi}\rangle_{tr}\right)
\right. \nonumber\\ &\otimes& \left. \left( \langle\tilde{\Phi}|_{tr}
\langle\tilde{\Phi}_{\perp}|_{tr} - \langle\tilde{\Phi}_{\perp}|_{tr}
\langle\tilde{\Phi}|_{tr} \right) \right\}. \nonumber
\end{eqnarray}

(ii) Overlap $\int \sqrt{Q_{\varphi}(n_{\varphi},n_{\varphi^{\perp}})
  \cdot Q_{{\varphi}^{\perp}}(n_{\varphi},n_{\varphi^{\perp}}) } d
n_{\varphi} d n_{\varphi^{\perp}}$ between discrete $Q_\varphi$-- and
$Q_{\varphi_\perp}$-- functions before preselection was calculated for
convolution of these functions with Gauss function.

(iii) The loophole observable is defined as
\begin{equation}
{\cal L} = {\cal P}_{n_{\varphi} < N_{\sigma}} \otimes {\cal
  P}_{n_{\varphi^{\perp}} < N_{\sigma}} + {\cal P}_{n_{\varphi} >
  N_{\sigma}} \otimes {\cal P}_{n_{\varphi^{\perp}} > N_{\sigma}}.
\label{Rh7a}
\end{equation}
It additively quantifies the violation of the preselection condition
corresponding to measurement of either of the two diagonal quadratures
in Fig.\ref{fig1}(b) where the discarded part of $Q$--functions is
located.

(iv) The differences in results in the Bell test obtained with
observables in Eq.(\ref{obs1}) and (\ref{obs2}) become more pronounced
and lead to contradicting conclusions for non-optimal thresholds
e.g. for $K_\mathrm{th}\!\!=\!\!1700$ and $N_{\sigma}\!\!=\!\!6000$,
$\left\langle \mathcal{B}\right\rangle_{\mathcal{A}}\!\!=\!\!-2.59>2$
and $\left\langle
\mathcal{B}\right\rangle_{\bar{\mathcal{A}}}\!\!=\!\!-1.90<2$.

(v) Table~\ref{table2} collects values of the correlation function
$\langle \mathcal{O}(0) \otimes \mathcal{O}(0) \rangle$, the Bell
parameter $\left\langle \mathcal{B}\right\rangle$ and the loophole
${\mathcal L}$.

(vi) Numerical simulations were performed with custom software written
in C++ and with use of Class Library for Numbers (CLN). CLN offers
greater precision over standard floating point numbers.

\acknowledgments

We thank A. Aiello, A. Lvovsky, M. Horodecki, M. Paw\l{}owski and P. Sekatski 
for discussions. This work was partially supported by UE IP projects QAP
and SCALA, Ministry of Science and Higher Education Grant
No. 2319/B/H03/2009/37 and 2619/B/H03/2010/38 and by the Foundation
for Polish Science.  Computation was performed in the TASK Center at
the Gda\'nsk University of Technology.

\begin{table}[p]
\begin{center}
\rotatebox{90}{
\begin{tabular}{|r|l|l|l|l|l|}
\hline
\multicolumn{1}{|c|}{\strut $K_\mathrm{th}$}&
\multicolumn{1}{|c|}{$N_\sigma=8200$}&
\multicolumn{1}{|c|}{$N_\sigma=8400$}&
\multicolumn{1}{|c|}{$N_\sigma=8600$}&
\multicolumn{1}{|c|}{$N_\sigma=8800$}\\
\hline\hline
\strut 1600&
$-0.996$\quad $-2.816$\quad $0.004$&
$-0.994$\quad $-2.810$\quad $0.006$&
$-0.990$\quad $-2.801$\quad $0.010$&
$-0.986$\quad $-2.788$\quad $0.014$\\
\strut &
$-0.983$\quad $-2.779$\quad $-$&
$-0.974$\quad $-2.756$\quad $-$&
$-0.961$\quad $-2.719$\quad $-$&
$-0.943$\quad $-2.668$\quad $-$\\
\strut 1700&
$-0.997$\quad $-2.819$\quad $0.003$&
$-0.997$\quad $-2.821$\quad $0.003$&
$-0.997$\quad $-2.820$\quad $0.003$&
$-0.996$\quad $-2.816$\quad $0.004$\\
\strut &
$-0.987$\quad $-2.791$\quad $-$&
$-0.990$\quad $-2.800$\quad $-$&
$-0.988$\quad $-2.795$\quad $-$&
$-0.982$\quad $-2.778$\quad $-$\\
\strut 1800&
$-0.992$\quad $-2.807$\quad $0.008$&
$-0.995$\quad $-2.815$\quad $0.005$&
$-0.997$\quad $-2.821$\quad $0.003$&
$-0.998$\quad $-2.824$\quad $0.002$\\
\strut &
$-0.970$\quad $-2.744$\quad $-$&
$-0.982$\quad $-2.777$\quad $-$&
$-0.989$\quad $-2.799$\quad $-$&
$-0.993$\quad $-2.809$\quad $-$\\
\strut 1900&
$-0.984$\quad $-2.784$\quad $0.016$&
$-0.989$\quad $-2.798$\quad $0.011$&
$-0.993$\quad $-2.809$\quad $0.007$&
$-0.996$\quad $-2.817$\quad $0.004$\\
\strut &
$-0.938$\quad $-2.653$\quad $-$&
$-0.957$\quad $-2.706$\quad $-$&
$-0.972$\quad $-2.749$\quad $-$&
$-0.984$\quad $-2.782$\quad $-$\\
\hline
\hline
\multicolumn{1}{|c|}{\strut $K_\mathrm{th}$}&
\multicolumn{1}{|c|}{$N_\sigma=5200$}&
\multicolumn{1}{|c|}{$N_\sigma=5400$}&
\multicolumn{1}{|c|}{$N_\sigma=5600$}&
\multicolumn{1}{|c|}{$N_\sigma=5800$}\\
\hline\hline
\strut 900&
$-0.973$\quad $-2.753$\quad $0.027$&
$-0.964$\quad $-2.728$\quad $0.036$&
$-0.952$\quad $-2.693$\quad $0.048$&
$-0.936$\quad $-2.649$\quad $0.064$\\
\strut &
$-0.894$\quad $-2.529$\quad $-$&
$-0.860$\quad $-2.434$\quad $-$&
$-0.814$\quad $-2.301$\quad $-$&
$-0.754$\quad $-2.132$\quad $-$\\
\strut 1000&
$-0.984$\quad $-2.783$\quad $0.016$&
$-0.983$\quad $-2.781$\quad $0.017$&
$-0.980$\quad $-2.771$\quad $0.020$&
$-0.973$\quad $-2.753$\quad $0.027$\\
\strut &
$-0.936$\quad $-2.648$\quad $-$&
$-0.934$\quad $-2.642$\quad $-$&
$-0.920$\quad $-2.603$\quad $-$&
$-0.895$\quad $-2.532$\quad $-$\\
\strut 1100&
$-0.979$\quad $-2.769$\quad $0.021$&
$-0.984$\quad $-2.784$\quad $0.016$&
$-0.987$\quad $-2.791$\quad $0.013$&
$-0.987$\quad $-2.792$\quad $0.013$\\
\strut &
$-0.918$\quad $-2.595$\quad $-$&
$-0.937$\quad $-2.651$\quad $-$&
$-0.948$\quad $-2.681$\quad $-$&
$-0.948$\quad $-2.683$\quad $-$\\
\strut 1200&
$-0.965$\quad $-2.730$\quad $0.035$&
$-0.974$\quad $-2.756$\quad $0.026$&
$-0.981$\quad $-2.776$\quad $0.019$&
$-0.986$\quad $-2.790$\quad $0.014$\\
\strut &
$-0.863$\quad $-2.442$\quad $-$&
$-0.899$\quad $-2.542$\quad $-$&
$-0.926$\quad $-2.619$\quad $-$&
$-0.945$\quad $-2.674$\quad $-$\\
\hline
\end{tabular}
}
\end{center}
\caption{ The correlation function, the Bell parameter $\langle {\cal
    B} \rangle $ and loophole ${\cal L}$ computed for the preselected
  ``macro-macro'' singlet with $m=10^3$ for selected $K_{th}$ and
  $N_{\sigma }$ using ${\cal A}(\varphi )$ -- the upper row and using
  $\bar{{\cal A}}(\varphi )$ -- the bottom row. The quantum character
  of a state is revealed by $2 < |\langle B \rangle| \leq 2
  \sqrt{2}$.}
\label{table2}
\end{table}

\end{document}